\documentclass[showpacs,prb,onecolumn]{revtex4}
\usepackage{amsfonts}
\usepackage{amsmath}
\usepackage{amssymb}
\usepackage{graphicx}

\begin{document}

\title{The Kagom\'{e} Antiferromagnet: A Schwinger-Boson Mean-Field Theory
Study}
\author{Peng Li, Haibin Su}
\affiliation{Division of Materials Science, Nanyang Technological University, 50 Nanyang
Avenue, Singapore 639798}
\author{Shun-Qing Shen}
\affiliation{Department of Physics, The University of Hong Kong, Pokfulam, Hong Kong,
China}

\begin{abstract}
The Heisenberg antiferromagnet on the Kagom\'{e} lattice is studied in the
framework of Schwinger-boson mean-field theory. Two solutions with different
symmetries are presented. One solution gives a conventional quantum state
with $\mathbf{q}=0$ order for all spin values. Another gives a gapped spin
liquid state for spin $S=1/2$ and a mixed state with both $\mathbf{q}=0$ and
$\sqrt{3}\times \sqrt{3}$ orders for spin $S>1/2$. We emphasize that the
mixed state exhibits two sets of peaks in the static spin structure factor.
And for the case of spin $S=1/2$, the gap value we obtained is consistent
with the previous numerical calculations by other means. We also discuss the
thermodynamic quantities such as the specific heat and magnetic
susceptibility at low temperatures and show that our result is in a good
agreement with the Mermin-Wagner theorem.
\end{abstract}

\date{\today }
\pacs{75.10.Jm, 75.30.Ds, 75.40.Cx}
\maketitle

\section{Introduction}

Two-dimensional (2D) geometrically frustrated Heisenberg antiferromagnets
(AFMs) are potential candiadates in the search of spin liquids
(spin-disordered states) from both theoretical and experimental
considerations. The first suggested candidate for spin liquid is the
triangular lattice with nearest-neighbor (NN) couplings, but unfortunately
it was finally revealed to exhibit $120^{\circ }$ spin long-range order by
extensive studies \cite{Anderson,Huse,Jolicoeur,Bernu,Capriotti}. People
began to resort to interactions beyond the NN coupling to realize the
spin-disordered state. Another potential candidate is the Kagom\'{e} AFM
\cite{Sachdev,Misguich}. It has been already known that there are two
possible ordered states in this system, the so-called $\mathbf{q}=0$ state
and $\sqrt{3}\times \sqrt{3}$ state \cite{Harris} (Fig. 1), while numerical
studies do not support any long range orders for the spin-1/2 system \cite%
{Zeng,Chalker}. This debate is still going on since numerical studies are
usually limited to finite lattices. It was also investigated extensively
whether the excitation spectra are gapless or not even if the system is
disordered. Numerical study with up to $36$ spins gives an estimation of the
energy gap smaller than $1/20$ of the exchange interaction \cite{Waldtmann}.
In a scenario of valence bond crystal with translational symmetry breaking,
the gap of the system is found to be very small \cite{Misguich,Syromyatnikov}%
. Experimentally, several Kagom\'{e}-like systems have been found \cite%
{Broholm}. One of the most intriguing experimental results is the large $%
T^{2}$ coefficient of the specific heat of spin-$\frac{3}{2}$ SrCr$_{9}$Ga$%
_{12}$O$_{19}$ \cite{Ramirez1990,Ramirez2000}, which suggests that a large
linear term exists in the density of states (DOS), $D(E)\sim \eta E$. A
numerical study of the spin-$\frac{1}{2}$ system suggests that the $T^{2}$
law of the specific heat can be inferred from the Heisenberg Hamiltonian
with the NN couplings at very low temperatures \cite{Sindzingre}. A
contractor renormalization calculation finds a columnar dimer order and
re-produces the $T^{2}$ specific heat for the spin-$\frac{1}{2}$ system, but
it still cannot tell whether it is gapless or not \cite{Budnik}. A very
recent projected wavefunction study suggests that the gapless mode may be
missed due to the limitation of finite lattice sites, and the gapless U($1$%
)-Dirac state produces the $T^{2}$ specific heat \cite{Ran}. So far many
aspects of the ground state remain to be mysteries.

\begin{figure}[tbp]
\begin{center}
\includegraphics[
height=1.6674in,
width=3.0165in
]{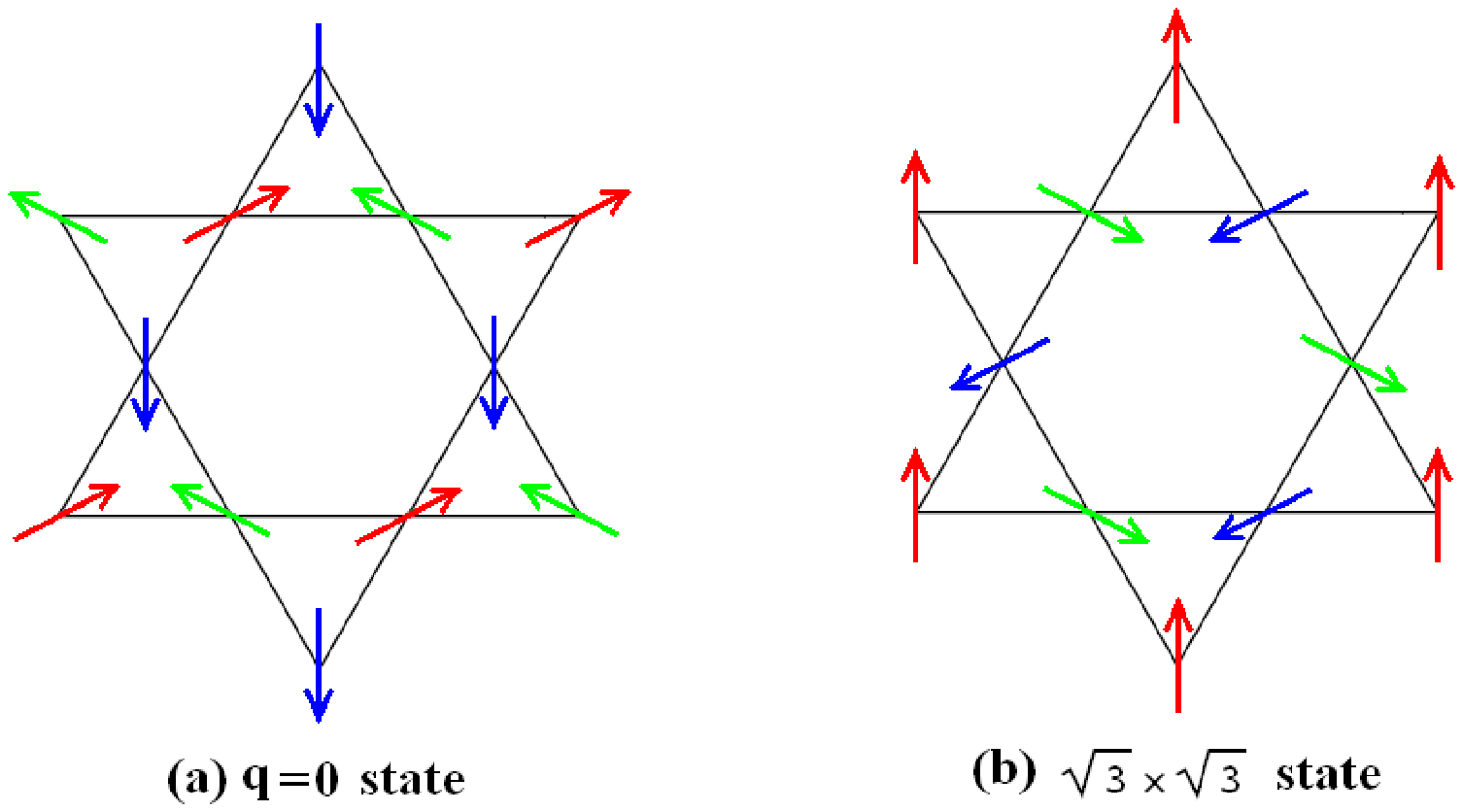}
\end{center}
\caption{(Color online) The $\mathbf{q}=0$ and $\protect\sqrt{3}\times
\protect\sqrt{3}$ ordered states.}
\end{figure}

In the theoretical aspect, it was known that the Schwinger-boson mean-field
theory (SBMFT) may provide a reliable description for both quantum ordered
and disordered antiferromagnets based on the picture of the resonating
valence bond (RVB) state \cite{Anderson87,Auerbach,Auerbachbook}. As a
merit, it does not prescribe any long-range order for the ground state in
advance, which should emerge naturally if the Schwinger bosons condense at
low temperatures. It was supposed that such a mean-field theory should be
reliable for large spins where quantum fluctuation is believed to be weak.
The theory has successfully captured the long-range order (LRO) of the
Heisenberg antiferromagnets on the square \cite{Auerbach,Auerbachbook} and
triangular \cite{Gazza,Shen} lattices at zero temperature, and is in
excellent agreement with the Mermin-Wagner theorem even for small spins. Of
course, it also has shortcomings, such as it predicts wrongly an energy gap
for a one-dimensional half-integer spin chains \cite{Auerbach,Auerbachbook}.
In previous works, SBMFT had been already applied to the Kagom\'{e} system.
Manuel \textit{et. al.} gave a Schwinger-boson approach to the $\mathbf{q}=0$
state and $\sqrt{3}\times \sqrt{3}$ state by including a third neighbour
interaction \cite{Manuel}. In a recent work by Wang, a new quantum
disordered state is proposed for the systems with physical spin values if
ring exchange interactions are introduced \cite{Wang}. In this paper, we
employ SBMFT to study the Heisenberg antiferromagnet with physical spin
values on the Kagom\'{e} lattice in a different approach and discover some
features of the system quantities. The gauge freedom due to the geometry
gives two solutions corresponding to two different types of the states \cite%
{Sachdev}. The first solution gives the $\mathbf{q}=0$ ordered state while
the second solution gives a mixed state with $\mathbf{q}=0$ order and $\sqrt{%
3}\times \sqrt{3}$ order for $S>\frac{1}{2}$, and a state with a very small
gap for $S=\frac{1}{2}$. It was shown that the strong quantum fluctuation of
quantum spin 1/2 may destroy the order states of higher spin and drive the
system to be disordered. The coexistence of two orders in a quantum state is
one of the key results in this paper. This result is revealed by the
detailed analysis of the static spin structure factors. For the ordered
states in both solutions, we show that the low-energy spectra for the
quasi-particles are linear in the momentum $\left\vert \mathbf{k}-\mathbf{k}%
^{\ast }\right\vert $ at the gapless point $\mathbf{k}^{\ast }$. As a
result, the density of state is linear in energy, the specific heat obeys
the $T^{2}$ law, and the uniform magnetic susceptibility is finite at zero
temperature.

The rest of the paper is arranged as follows. The general formalism of the
Schwinger-boson mean field theory is presented in Sec. II. We introduce two
types of mean field parameters, and expect to capture the key features of
quantum spin state on the Kagome lattice. A set of mean field equation is
established by means of the Matsubara Green function techniques. In Sec.
III, the numerical solutions of the mean field equations are given. We focus
on the ground state properties for the system and show that the ground state
of spin 1/2 has a finite energy gap to the excited states and is spin
disordered while for larger spin the ground state is spin ordered. Finally,
a brief discussion is presented in Sec. IV.

\begin{figure}[tbp]
\begin{center}
\includegraphics[
height=1.6903in,
width=3.281in
]{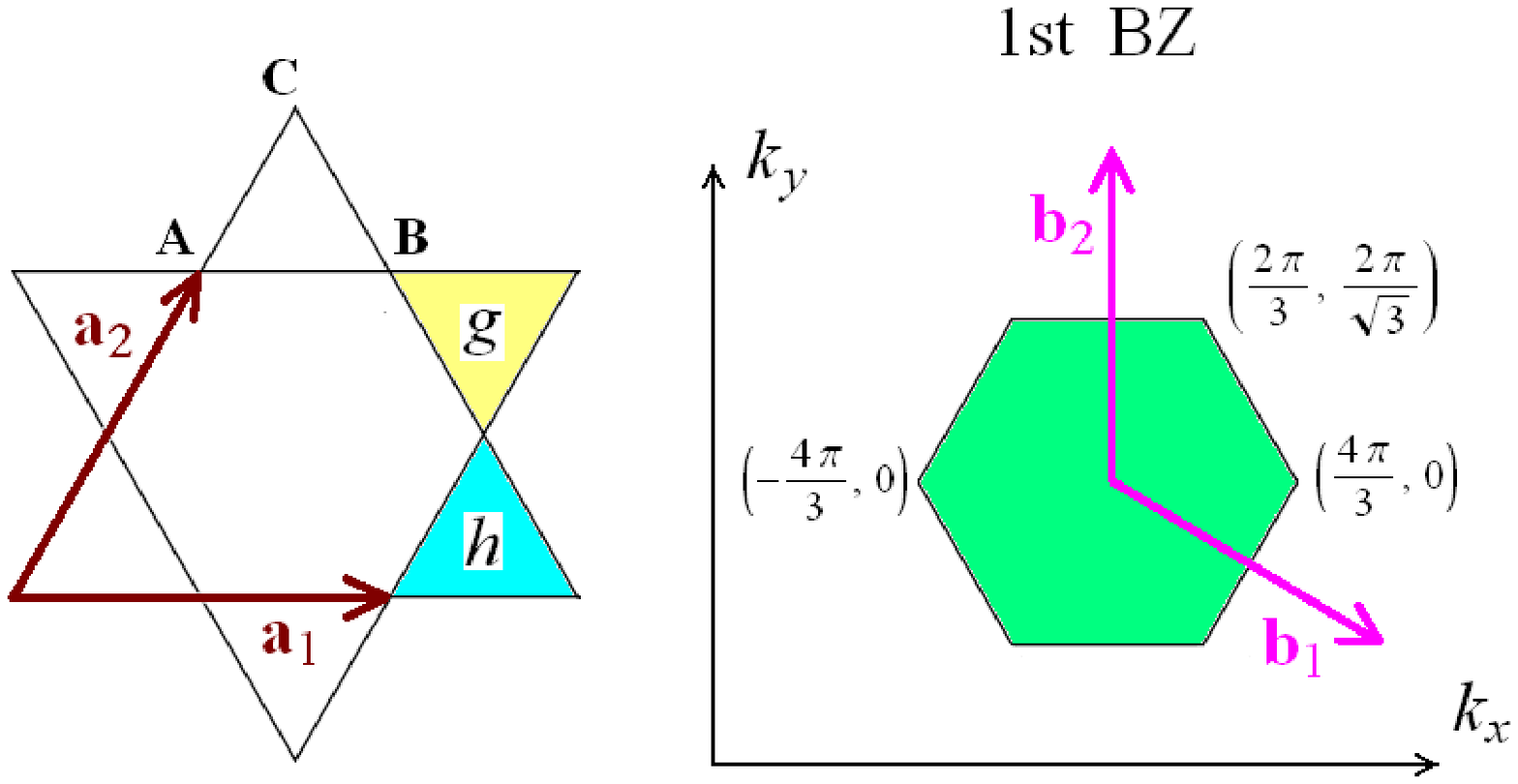}
\end{center}
\caption{(Color online) The primitive cell and the first Brillouin zone
(with an area of $A_{BZ}=8\protect\pi ^{2}/\protect\sqrt{3}$) of the Kagom%
\'{e} lattice. The primitive translation vectors of the direct and
reciprocal lattices are $\left( \mathbf{a}_{1}=(1,0),\mathbf{a}_{2}=(1/2,%
\protect\sqrt{3}/2)\right) $ and $\left( \mathbf{b}_{1}=\left( 2\protect\pi %
,-2\protect\pi /\protect\sqrt{3}\right) ,\mathbf{b}_{2}=\left( 0,4\protect%
\pi /\protect\sqrt{3}\right) \right) ,$ respectively.}
\end{figure}

\section{Schwinger-boson mean-field theory}

We start with the Heisenberg Hamiltonian on the Kagom\'{e} lattice,%
\begin{equation}
H=J\sum_{\left\langle i,m;j,m^{\prime }\right\rangle }\mathbf{S}_{i,m}\cdot
\mathbf{S}_{j,m^{\prime }},
\end{equation}%
where $i$ and $j$ are indices of the periodic Bravis lattice, $m$ and $%
m^{\prime }$ are sublattice indices, $A,B,$or $C$ as indicated in Fig. 2,
and the notation $\left\langle i,m;j,m^{\prime }\right\rangle $ means all
possible NN pairs of lattice sites. The exchange interaction will be set as
the unit of energy, $J=1$. Note that the lattice constant is double of the
triangle parameter $l_{0}$ and we set $2l_{0}=1$ for simplicity. We choose
the primitive cell and the first Brillouin zone as in Fig. 2. In the
framework of Schwinger boson theory \cite{Auerbach}, a pair of hard-core
bosons is introduced to represent one quantum spin $S$ at each site,
\begin{equation}
S_{i,m}^{+}=b_{i,m,\uparrow }^{\dag }b_{i,m,\downarrow
},S_{i,m}^{-}=b_{i,m,\downarrow }^{\dag }b_{i,m,\uparrow },S_{i,m}^{z}=\frac{%
1}{2}\left( b_{i,m,\uparrow }^{\dag }b_{i,m,\uparrow }-b_{i,m,\downarrow
}^{\dag }b_{i,m,\downarrow }\right) ,
\end{equation}%
with the constraint, $b_{i,m,\uparrow }^{\dag }b_{i,m,\uparrow
}+b_{i,m,\downarrow }^{\dag }b_{i,m,\downarrow }\equiv 2S.$ In this
representation, the Hamiltonian can be expressed as%
\begin{equation}
H=-\sum_{\left\langle i,m;j,m^{\prime }\right\rangle }\left( 2\Delta
_{i,m;j,m^{\prime }}^{\dag }\cdot \Delta _{i,m;j,m^{\prime }}+S^{2}\right)
+\sum_{i,m}2\lambda _{i,m}(b_{i,m,\uparrow }^{\dag }b_{i,m,\uparrow
}+b_{i,m,\downarrow }^{\dag }b_{i,m,\downarrow }-2S),
\end{equation}%
where $\Delta _{i,m;j,m^{\prime }}\equiv \frac{1}{2}(b_{i,m,\uparrow
}b_{j,m^{\prime },\downarrow }-b_{i,m,\downarrow }b_{j,m^{\prime },\uparrow
})$ and $N_{\Lambda }$ is the number of primitive cells (i.e. the total
number of lattice sites is $3N_{\Lambda }$). The Lagrange multipliers $%
\lambda _{i,m}$ is introduced to realize the constraints of the number of
Schwinger bosons at each site. Due to translational symmetry, we will set $%
\lambda _{i,m}=\lambda $ to simplify the problem. In the mean field
approach, one can introduce the mean-field parameter $\left\langle \Delta
_{i,m;j,m^{\prime }}\right\rangle $, and decompose $\Delta _{i,m;j,m^{\prime
}}^{\dag }\cdot \Delta _{i,m;j,m^{\prime }}$ into $\left\langle \Delta
_{i,m;j,m^{\prime }}^{\dag }\right\rangle \Delta _{i,m;j,m^{\prime }}+\Delta
_{i,m;j,m^{\prime }}^{\dag }\left\langle \Delta i,m;j,m^{\prime
}\right\rangle -\left\vert \Delta _{i,m;j,m^{\prime }}\right\vert ^{2}$
where $\left\langle \cdots \right\rangle $ represents the thermodynamic
average of the physical quantity. This procedure can also be formulated
equivalently as the Hubbard-Stratanovich transformation \cite{Auerbachbook}.
In a suitable gauge, the bond parameter $\left\langle \Delta
_{i,m;j,m^{\prime }}\right\rangle $ can be taken to be real \cite{Read}.
Notice there are two different solutions that can be obtained by the
relation of the mean fields on the two adjacent triangles (labeled as $g$
and $h$ in Fig. 2). Following the spirit of Sachdev's discussion on gauge
freedom of the mean-field parameter \cite{Sachdev}, we present two schemes:
(i) $\Delta =\left\langle \Delta _{i,m;j,m^{\prime }}^{g}\right\rangle
=\left\langle \Delta _{i,m;j,m^{\prime }}^{h}\right\rangle $; (ii) $\Delta
=\left\langle \Delta _{i,m;j,m^{\prime }}^{g}\right\rangle =-\left\langle
\Delta _{i,m;j,m^{\prime }}^{h}\right\rangle $. They produce two physically
distinct states that cannot be transformed into each other by gauge
transformations.

In both schemes the effective Hamiltonian is decomposed into the quadratic
form of $b_{i,m,\uparrow }$ and $b_{i,m,\downarrow }$, and the mean-field
theories for the two schemes have almost the same formalism. In the
following deduction, we shall point out their differences where it is
appropriate. We can introduce three pairs of $b_{\mathbf{k},m,\mu }^{\dag }$
and $b_{\mathbf{k},m,\mu }$ and $\mu =\uparrow $ or $\downarrow $ in the
Fourier transform such that the effective Hamiltonian can be written in a
compact form with the help of Kronecker product,
\begin{subequations}
\begin{align}
H_{eff}& =\sum_{\mathbf{k}}\Psi _{\mathbf{k}}^{\dag }H_{mf}\Psi _{\mathbf{k}%
}+\varepsilon _{0}, \\
H_{mf}& =\lambda I_{0}+\Delta \sigma _{x}\otimes \left[ \gamma _{1}\Omega
_{z}-\gamma _{2}\Omega _{y}+\gamma _{3}\Omega _{x}\right] \otimes \sigma
_{y}, \\
\varepsilon _{0}& =12N_{\Lambda }\Delta ^{2}-6\lambda N_{\Lambda }\left(
2S+1\right) +6N_{\Lambda }S^{2},
\end{align}%
where the Nambu spinor is introduced
\end{subequations}
\begin{equation}
\Psi _{\mathbf{k}}^{\dag }=(\psi _{\mathbf{k}}^{\dag },\psi _{-\mathbf{k}}),%
\text{ }\psi _{\mathbf{k}}^{\dag }=(b_{\mathbf{k},A,\uparrow }^{\dag }b_{%
\mathbf{k},A,\downarrow }^{\dag }b_{\mathbf{k},B,\uparrow }^{\dag }b_{%
\mathbf{k},B,\downarrow }^{\dag }b_{\mathbf{k},C,\uparrow }^{\dag }b_{%
\mathbf{k},C,\downarrow }^{\dag }),  \label{Nambu}
\end{equation}%
$I_{0}$ is a $12\times 12$ unit matrix, $\sigma _{\alpha }\left( \alpha
=x,y,z\right) $ are $2\times 2$ Pauli matrices, $\Omega _{\alpha }\left(
\alpha =x,y,z\right) $ are $3\times 3$ Hermitian matrices
\begin{equation}
\Omega _{x}=\left(
\begin{array}{ccc}
0 & 0 & 0 \\
0 & 0 & -i \\
0 & i & 0%
\end{array}%
\right) ,\Omega _{y}=\left(
\begin{array}{ccc}
0 & 0 & i \\
0 & 0 & 0 \\
-i & 0 & 0%
\end{array}%
\right) ,\Omega _{z}=\left(
\begin{array}{ccc}
0 & -i & 0 \\
i & 0 & 0 \\
0 & 0 & 0%
\end{array}%
\right) ,
\end{equation}%
and for the two schemes we have (i) $\gamma _{1}=\cos \frac{k_{x}}{2},\gamma
_{2}=\cos \frac{k_{x}+\sqrt{3}k_{y}}{4},\gamma _{3}=\cos \frac{k_{x}-\sqrt{3}%
k_{y}}{4}$; and (ii) $\gamma _{1}=\sin \frac{k_{x}}{2},\gamma _{2}=\sin
\frac{k_{x}+\sqrt{3}k_{y}}{4},\gamma _{3}=\sin \frac{k_{x}-\sqrt{3}k_{y}}{4}$%
, respectively.

Let us define the Matsubara Green's function (a $12\times 12$ matrix) by the
outer product of the Nambu spinor Eq. (\ref{Nambu}),%
\begin{equation}
G\left( \mathbf{k},\tau \right) =-\left\langle T_{\tau }\Psi _{\mathbf{k}%
}\left( \tau \right) \Psi _{\mathbf{k}}^{\dag }\right\rangle ,
\end{equation}%
where $\tau $ is the imaginary time and $\Psi _{\mathbf{k}}\left( \tau
\right) =e^{\tau H_{eff}}\Psi _{\mathbf{k}}e^{-\tau H_{eff}}$. Then physical
quantities concerning the average of two operators can be readily expressed
by the matrix elements, e.g. $\left\langle b_{\mathbf{k},1,\uparrow }b_{%
\mathbf{k},1,\uparrow }^{\dag }\right\rangle =-G_{1,1}\left( \mathbf{k},\tau
=0^{+}\right) $. And the physcial quantities concerning the average of four
operators, such as the correlation functions, can be decomposed into the
averages of two operators through the Wick theorem. We shall use these facts
to establish the mean-field equations and the static spin structure factors
later.

It is easy to prove that the Matsubara Green's function in Matsubara
frequency $\omega _{n}=2n\pi /\beta $ ($n$ is an integer for bosons) can be
worked out by (also a $12\times 12$ matrix)
\begin{equation}
G(\mathbf{k},i\omega _{n})=\left[ i\omega _{n}\sigma _{z}\otimes \Omega
_{0}\otimes \sigma _{0}-H_{mf}\right] ^{-1},
\end{equation}%
where $\Omega _{0}$ and $\sigma _{0}$ are $3\times 3$ and $2\times 2$ unit
matrices. Because of symmetry, some elements are the same. Fortunately,
these functions can be calculated analytically, and the lengthy expressions
will be presented elsewhere. From the poles of single particle Matsubara
function, the six branches of the energy spectra of quasi-particles can be
read out in the first Brillouin zone. Two are the flat bands $\omega _{1,\mu
}=\lambda $ and other four-fold degenerate bands are $\omega _{2,\mu }\left(
\mathbf{k}\right) =\omega _{3,\mu }\left( \mathbf{k}\right) =\omega \left(
\mathbf{k}\right) $ with $\omega \left( \mathbf{k}\right) =\sqrt{\lambda
^{2}-\Delta ^{2}\gamma ^{2}\left( \mathbf{k}\right) }$ with $\gamma
^{2}\left( \mathbf{k}\right) =\gamma _{1}^{2}+\gamma _{2}^{2}+\gamma
_{3}^{2} $. The mean-field parameter $\Delta $ and the Lagrangian multiplier
$\lambda $ should be determined self-consistently. The mean field can be
evaluated by reading the elements of the Matsubara Green function matrix
after the Fourier transformation, e.g.
\begin{eqnarray}
\Delta &=&\frac{1}{2}\left( \left\langle b_{i,A,\uparrow }b_{i,B,\downarrow
}\right\rangle -\left\langle b_{i,A,\downarrow }b_{i,B,\uparrow
}\right\rangle \right)  \notag \\
&=&\frac{-1}{2\beta N_{\Lambda }}\sum_{\mathbf{k,}i\omega
_{n}}e^{-ik_{x}/2-i\omega _{n}0^{+}}\left[ G_{10,1}\left( \mathbf{k},i\omega
_{n}\right) +G_{9,2}\left( \mathbf{k},i\omega _{n}\right) \right] .
\end{eqnarray}%
Another constraint is that we should use the average value in the
thermodynamic limit $\left\langle b_{i,m,\uparrow }^{\dag }b_{i,m,\uparrow
}+b_{i,m,\downarrow }^{\dag }b_{i,m,\downarrow }\right\rangle =2S$ to\
replace the original constraint. These two facts lead to a set of the
mean-field equations for $\Delta $ and $\lambda $,
\begin{subequations}
\label{MFE}
\begin{align}
3S+1& =2n_{B}\left( \lambda \right) +\frac{1}{N_{\Lambda }}\sum_{\mathbf{k}}%
\frac{\lambda }{\omega \left( \mathbf{k}\right) }\left[ 1+2n_{B}\left(
\omega \left( \mathbf{k}\right) \right) \right] ,  \label{MFEa} \\
\Delta & =\frac{1}{6}\frac{1}{N_{\Lambda }}\sum_{\mathbf{k}}\frac{\Delta
\gamma ^{2}\left( \mathbf{k}\right) }{\omega \left( \mathbf{k}\right) }\left[
1+2n_{B}\left( \omega \left( \mathbf{k}\right) \right) \right] ,
\end{align}%
where $n_{B}\left( \omega \left( \mathbf{k}\right) \right) =\left[ e^{\omega
\left( \mathbf{k}\right) /T}-1\right] ^{-1}$ is the Bose-Einstein
distribution function with temperature $T$. In the thermodynamical limit $%
N_{\Lambda }\rightarrow \infty $, the momentum sum is replaced by the
integral over the first Brillouin zone (Fig. 2), $\frac{1}{N_{\Lambda }}%
\sum_{\mathbf{k}}\rightarrow \int \frac{d^{2}k}{A_{BZ}}$, where $A_{BZ}=%
\frac{8\pi ^{2}}{\sqrt{3}}$ is the area of the first Brillouin zone. When
the Schwinger bosons condensation occurs, i.e. the solution gives a gapless
spectrum $\omega \left( \mathbf{k}^{\ast }\right) =0$, we can extract a
condensation term \cite{Auerbachbook} in the momentum summation of the first
equation, Eq. (\ref{MFEa}),
\end{subequations}
\begin{equation}
\rho _{0}(T)=\frac{\lambda }{N_{\Lambda }\omega \left( \mathbf{k}^{\ast
}\right) }\left[ 1+2n_{B}\left( \omega \left( \mathbf{k}^{\ast }\right)
\right) \right] .  \label{condensation}
\end{equation}%
With the help of the mean-field equations at zero temperature, we can obtain
the simplified form of ground energy per bond
\begin{equation}
E_{0}/6N_{\Lambda }=-2\Delta ^{2}+S^{2}.  \label{E0}
\end{equation}

\section{Solutions}

To solve the mean-field equations, let us introduce dimensionless
quantities, $\widetilde{\Delta }=\frac{\Delta }{\lambda }$ and $\widetilde{T}%
=\frac{T}{\lambda }$. Then the mean-field equations become,
\begin{subequations}
\begin{align}
3S+1& =\coth \left( \frac{1}{2\widetilde{T}}\right) -1+\rho _{0}(\widetilde{T%
})+I_{0}(\widetilde{T}), \\
\Delta & =\frac{\widetilde{\Delta }\gamma ^{2}\left( \mathbf{k}^{\ast
}\right) }{6}\rho _{0}(\widetilde{T})+I_{1}(\widetilde{T}),
\end{align}%
with the definitions of two integrals
\end{subequations}
\begin{subequations}
\label{inte}
\begin{eqnarray}
I_{0}(\widetilde{T}) &=&\int \frac{d^{2}k}{A_{BZ}}\frac{1}{\sqrt{1-%
\widetilde{\Delta }^{2}\gamma ^{2}\left( \mathbf{k}\right) }}\coth \frac{%
\sqrt{1-\widetilde{\Delta }^{2}\gamma ^{2}\left( \mathbf{k}\right) }}{2%
\widetilde{T}}, \\
I_{1}(\widetilde{T}) &=&\frac{1}{6}\int \frac{d^{2}k}{A_{BZ}}\frac{%
\widetilde{\Delta }\gamma ^{2}\left( \mathbf{k}\right) }{\sqrt{1-\widetilde{%
\Delta }^{2}\gamma ^{2}\left( \mathbf{k}\right) }}\coth \frac{\sqrt{1-%
\widetilde{\Delta }^{2}\gamma ^{2}\left( \mathbf{k}\right) }}{2\widetilde{T}}%
.
\end{eqnarray}

First, we solve the equations at zero temperature. For the scheme (i), we
found the first integral is bounded from above by the value $I_{0}=1.75097$
at $\widetilde{\Delta }=\frac{1}{\sqrt{3}}$. We notice that a nonzero
condensation density persists as long as $S\geq S_{c}\approx 0.25032$.
Numerical solutions are $\rho _{0}=3S-0.75097,$ $\Delta =\frac{\sqrt{3}}{2}%
S+0.091132,$ and $\lambda =\frac{3}{2}S+0.15785$. The condensation occurs at
the $\Gamma $ point, $\mathbf{k}^{\ast }=\left( 0,0\right) $. For the scheme
(ii) we find that the first integral is bounded from above by the value $%
I_{0}=2.68932$ at $\widetilde{\Delta }=\frac{2}{3}$ and the critical spin
value is $S_{c}\approx 0.5631$. Thus for $S=\frac{1}{2}$, we obtain a spin
liquid because the spectrum is gapped with numerical solution $\rho _{0}=0,$
$\Delta =0.493757,$ and $\lambda =0.741905$. The value of the gap is $\Delta
_{gap}=0.0434$, which is coincident with the recent numerical estimation
that the gap is smaller than $\frac{1}{20}$ \cite{Waldtmann}. For larger
spins $S>\frac{1}{2}$, numerical solutions are $\rho _{0}=3S-1.68932,$ $%
\Delta =\frac{3}{4}S+0.118784,$ and $\lambda =\frac{9}{8}S+0.178176$. The
condensation occurs at six corner $K$ points, e.g. $\mathbf{k}^{\ast
}=\left( \frac{4\pi }{3},0\right) $.

Now, we discuss the asymptotic behavior of the solution near zero
temperature. Our numerical results show that the condensation only occurs at
zero temperature, which is in agreement with the Mermin-Wagner theorem for
two-dimensional Heisenberg systems. Numerical calculations tell us that the
gapless solution can only exist at zero temperature and an energy gap opens
at finite temperature, which behaves as $\Delta _{gap}\varpropto e^{-c/T}$
as $T\rightarrow 0$. We give an example for the case of $S=\frac{1}{2}$ in
the scheme (i). The curve for $\ln \Delta _{gap}\sim \frac{1}{T}$ is plotted
in Fig. 3. The lowest finite temperature we approach is $T\approx 0.03177$
(in unit of coupling $J$), where we get a small gap $\Delta _{gap}\approx
1.944\times 10^{-6}$.

\begin{figure}[tbp]
\begin{center}
\includegraphics[
trim=0.000000in 0.000000in 0.003146in -0.009125in,
height=4.99cm,
width=7.98cm
]{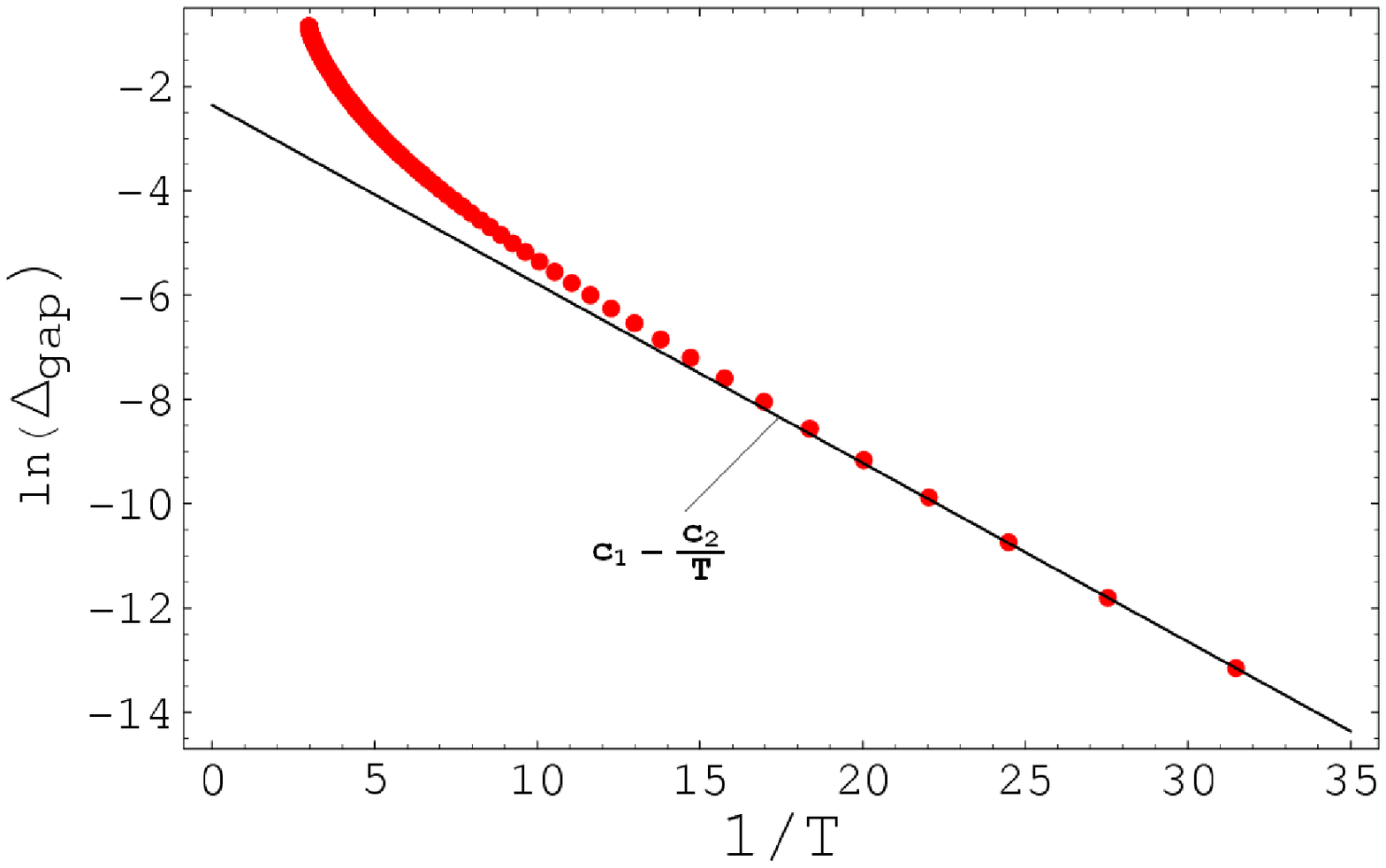}
\end{center}
\caption{(Color online) The asymptotic behavior of the gap near zero
temperature for the case of $S=1/2$ in scheme (i). The dots are numerical
solutions. The solid line is a linear fit with $c_{1}\approx -2.36$ and $%
c_{2}\approx 0.34$, which gives $\Delta _{gap}\approx 0.094e^{-0.34/T}$ as $%
T\rightarrow 0$.}
\end{figure}

\begin{figure}[tbp]
\begin{center}
\includegraphics[
trim=0.000000in 0.000000in 0.003146in -0.009125in,
height=6.8281cm,
width=7.9774cm
]{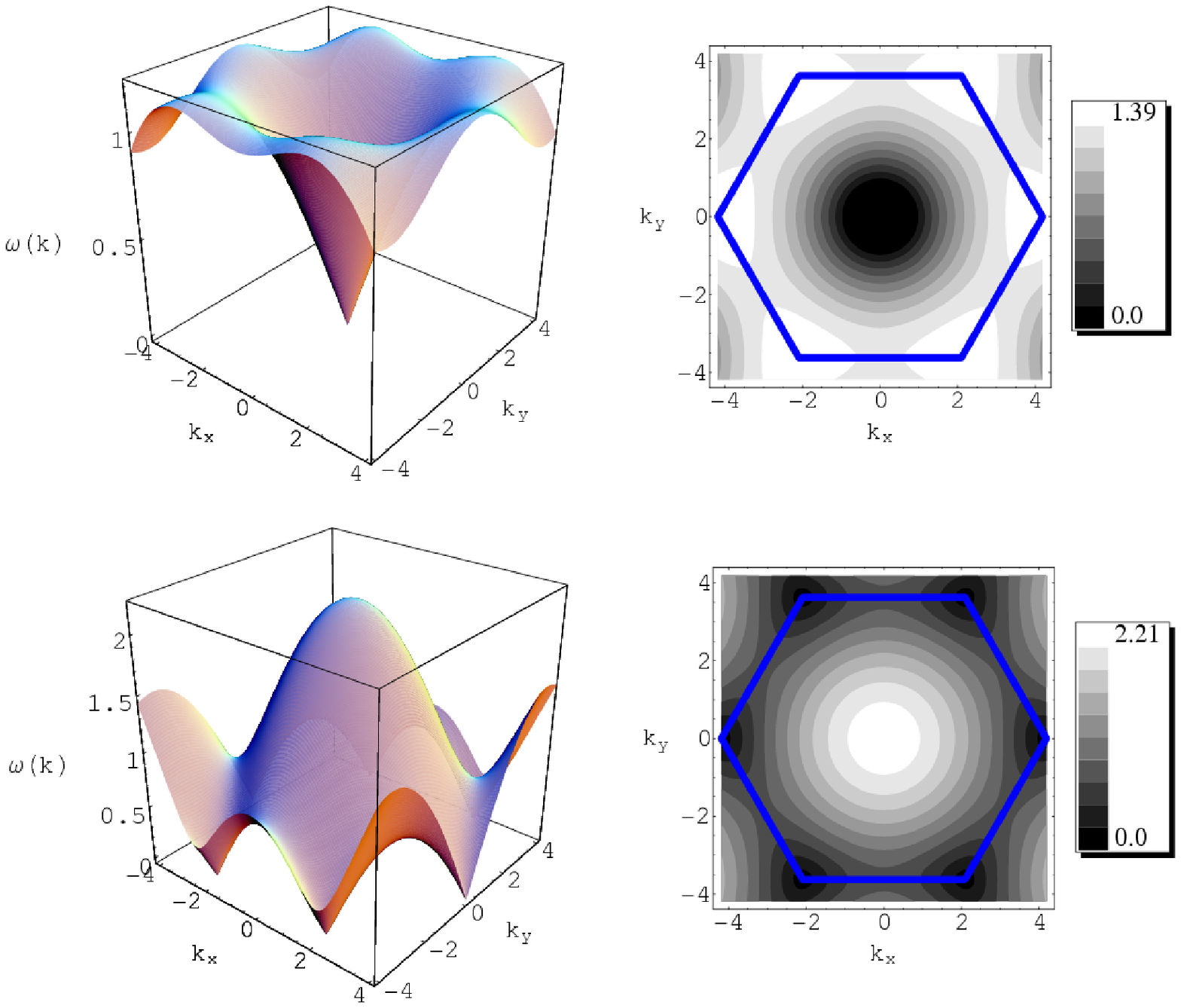}
\end{center}
\caption{(Color online) The cone-shaped quasiparticle spectra for the
gapless solutions. Here we set $S=3/2$. The upper spectrum is for scheme
(i). The lower one is for scheme (ii). The interior area of the blue hexagon
is the first Brillouin zone.}
\end{figure}

In the following, we turn to the relevant thermodynamical quantities and the
patterns of LRO at zero temperature. The gapless spectra for both ordered
states are exemplified in Fig. 4. It is clear that the cone-shaped spectra
are linear in the momentum
\end{subequations}
\begin{equation}
\omega \left( \mathbf{k}\right) \approx \alpha \left\vert \mathbf{k}-\mathbf{%
k}^{\ast }\right\vert ,
\end{equation}%
where $\alpha =\frac{\lambda }{2\sqrt{2}}$ for the scheme (i) and $\alpha =%
\frac{\lambda }{2\sqrt{3}}\ $for the scheme (ii). Correspondingly, the
density of state (DOS) of the quasiparticles linear in the low energy,
\begin{equation}
D\left( E\right) \approx \eta E+O(E^{3}),
\end{equation}%
where $\eta =\frac{8\sqrt{3}}{\pi \lambda ^{2}}$ for the scheme (i) and $%
\eta =\frac{24}{\pi \lambda ^{2}}\ $for the scheme (ii). As a result, the $%
T^{2}$ law of specific heat is anticipated for both ordered states%
\begin{equation}
C_{V}/N_{\Lambda }\sim 6\zeta \left( 3\right) \eta \left( \frac{T}{J}\right)
^{2}  \label{Cv}
\end{equation}%
at very low temperatures.

\begin{figure}[tbp]
\begin{center}
\includegraphics[
height=3.599in,
width=3.3972in
]{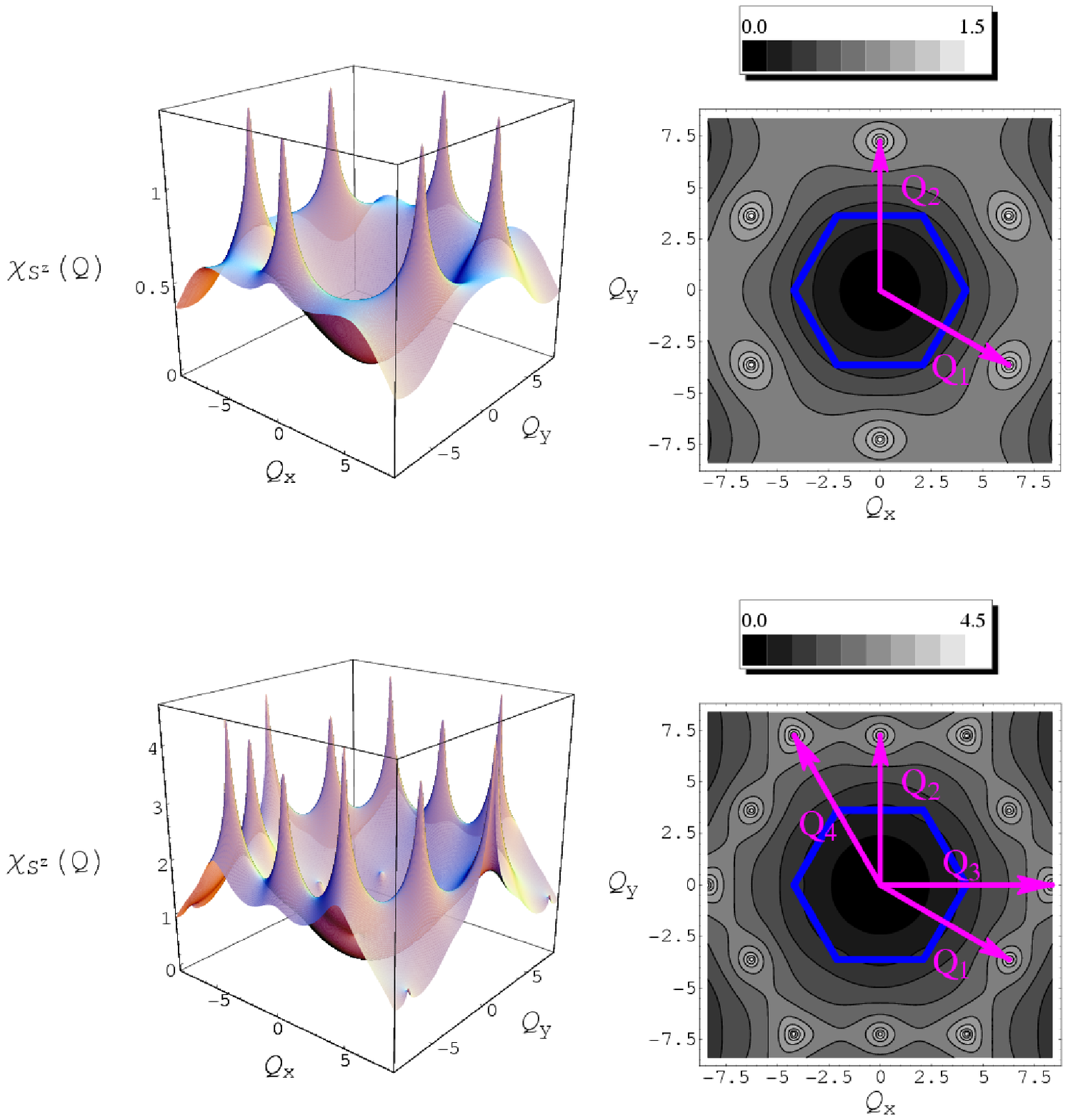}
\end{center}
\caption{(Color online) The spin structure factors $\protect\chi %
_{S^{z}}\left( \mathbf{Q}\right) $ for the two mean-field schemes in the
text. The upper one is for scheme (i), and the lower one is for scheme (ii).
Here we set $S=3/2$. The divergent peaks signal the existence of LRO. Other
spin values have similar results, except for the case of $S=1/2$ in scheme
(ii), where the peaks are of finite height because the solution is gapped.
The interior area of the blue hexagon is the first Brillouin zone.}
\end{figure}

The non-zero value of $\rho _{0}$ means the condensation of the
quasi-particles and the existence of LRO at zero-temperature. To disclose
the ordered patterns of the ground states we need to calculate the static
spin structure factor
\begin{equation}
\chi _{S^{z}}\left( \mathbf{Q}\right) =\lim_{\tau \rightarrow 0}\left\langle
T_{\tau }S_{\mathbf{Q}}^{z}(\tau )S_{-\mathbf{Q}}^{z}\right\rangle ,
\label{sus}
\end{equation}%
where $\tau $ is the imaginary time and $S_{\mathbf{Q}}^{z}=\frac{1}{\sqrt{%
N_{\Lambda }}}\sum_{i,m}S_{i,m}^{z}e^{i\mathbf{Q}\cdot \mathbf{r}_{i,m}}$. A
detailed calculation shows that the static spin structure factor is
isotropic, $\left\langle S_{\mathbf{Q}}^{x}S_{-\mathbf{Q}}^{x}\right\rangle
=\left\langle S_{\mathbf{Q}}^{y}S_{-\mathbf{Q}}^{y}\right\rangle
=\left\langle S_{\mathbf{Q}}^{z}S_{-\mathbf{Q}}^{z}\right\rangle ,$ which
indicates that the ground state is invariant under the spin rotation.
Analytically, we have that the total spin $\left\langle \mathbf{S}%
_{tot}^{2}\right\rangle =3N_{\Lambda }\chi _{S^{z}}\left( 0\right) =0$ for
both ordered states, which is consistent with the consequence of exact
diagonalization techniques for \cite{Zeng,Waldtmann}. The divergent peaks of
$\chi _{S^{z}}\left( \mathbf{Q}\right) $\ signal the existence of LRO as
shown in Fig. 5. (Notice that for the case of $S=1/2$ in the scheme (ii),
the peaks are of finite height because the solution is gapped. We do not
show it here.) Careful calculation shows the value of the divergent peak is
proportional to the number of primitive cells%
\begin{equation}
\chi _{S^{z}}\left( \mathbf{Q}^{\ast }\right) \propto \frac{\lambda ^{2}%
\left[ 1+2n_{B}\left( \omega \left( \mathbf{k}^{\ast }\right) \right) \right]
\left[ 1+2n_{B}\left( \omega \left( \mathbf{k}^{\ast }+\mathbf{Q}^{\ast
}\right) \right) \right] }{N_{\Lambda }\omega \left( \mathbf{k}^{\ast
}\right) \omega \left( \mathbf{k}^{\ast }+\mathbf{Q}^{\ast }\right) }%
=N_{\Lambda }\rho _{0}^{2}.
\end{equation}%
$\propto $Notice the primitive cell contains more than one site, the
replicative area of $\chi _{S^{z}}\left( \mathbf{Q}\right) $ is $4$ times of
the area of the first Brillouin zone $A_{BZ}$ (Fig. 5). This can be easily
verified by the definition, Eq. (\ref{sus}).

For the scheme (i), a characteristic feature of the static structure factor $%
\chi _{S^{z}}\left( \mathbf{Q}\right) $ is the six divergent peaks, which
are located at the wave vectors, $\mathbf{Q}^{\ast }\in \left\{ \pm \mathbf{Q%
}_{1},\pm \mathbf{Q}_{2},\pm \left( \mathbf{Q}_{1}+\mathbf{Q}_{2}\right)
\right\} $ with $\mathbf{Q}_{1}=\mathbf{b}_{1}$ and $\mathbf{Q}_{2}=\mathbf{b%
}_{2}$. At the divergent peaks, say e.g. at $\mathbf{Q}=\pm \mathbf{Q}_{1}$,
one gets $S_{\pm \mathbf{Q}_{1}}^{z}=\frac{1}{\sqrt{N_{\Lambda }}}%
\sum_{i}\left( S_{i,A}^{z}+e^{\pm i\pi }S_{i,B}^{z}+S_{i,C}^{z}\right) $.
Combining it with the facts that, $\chi _{S^{z}}\left( 0\right) =0$, we draw
a conclusion that the configuration of LRO has the $\mathbf{q}=0$ order. The
$\mathbf{q}=0$ order is marked by the neutron scattering peak at the distance%
\begin{equation}
\left\vert \mathbf{Q}_{1}\right\vert =\frac{4\pi }{\sqrt{3}}.
\end{equation}%
should mark the neutron scattering peak position.

For the scheme (ii), a similar analysis leads to another type of order
pattern. In this case the divergent peaks are located at the wave vectors, $%
\mathbf{Q}^{\ast }\in \left\{ \pm \mathbf{Q}_{1},\pm \mathbf{Q}_{2},\pm
\left( \mathbf{Q}_{1}+\mathbf{Q}_{2}\right) ,\pm \mathbf{Q}_{3},\pm \mathbf{Q%
}_{4},\pm \left( \mathbf{Q}_{3}+\mathbf{Q}_{4}\right) \right\} $ with $%
\mathbf{Q}_{3}=\frac{2}{3}\left( 2\mathbf{b}_{1}+\mathbf{b}_{2}\right) $ and
$\mathbf{Q}_{4}=\frac{2}{3}\left( -\mathbf{b}_{1}+\mathbf{b}_{2}\right) $.
The patterns corresponding $\mathbf{Q}_{3}$ and $\mathbf{Q}_{4}$ peaks are
the usual $\sqrt{3}\times \sqrt{3}$ order (Fig. 1). So for the scheme (ii)
we obtain a solution with mixed $\mathbf{q}=0$ order and $\sqrt{3}\times
\sqrt{3}$ order. Notice the mixture of two orders originate from quantum
mechanical superposition. This may be an exotic state with coexistence of
two distinct orders. The $\sqrt{3}\times \sqrt{3}$ order is marked by
neutron scattering peak at the distance%
\begin{equation}
\left\vert \mathbf{Q}_{3}\right\vert =\frac{8\pi }{3}.
\end{equation}%
Our calculation found the peak at $\left\vert \mathbf{Q}_{3}\right\vert $ is
a little stronger than that at $\left\vert \mathbf{Q}_{1}\right\vert $,
since the ratio of the two sets of divergent peaks is%
\begin{equation}
\frac{\chi _{S^{z}}\left( \mathbf{Q}_{3}\right) }{\chi _{S^{z}}\left(
\mathbf{Q}_{1}\right) }=\frac{3}{2}.
\end{equation}

The uniform magnetic susceptibility can be obtained by the analytic
continuation%
\begin{align}
\chi _{M}& =\lim_{\mathbf{Q}\rightarrow 0}\lim_{i\omega _{n}\rightarrow
0}\chi _{S^{z}}\left( \mathbf{Q},i\omega _{n}\right)  \notag \\
& =\int \frac{d^{2}k}{A_{BZ}}\frac{\Delta ^{4}\left[ \gamma _{1}^{2}\gamma
_{2}^{2}+\gamma _{2}^{2}\gamma _{3}^{2}+\gamma _{3}^{2}\gamma _{1}^{2}\right]
}{2\omega \left( \mathbf{k}\right) \left[ \lambda -\omega \left( \mathbf{k}%
\right) \right] \left[ \lambda +\omega \left( \mathbf{k}\right) \right] ^{3}}%
.
\end{align}%
$\chi _{M}$ has a finite value at zero temperature since the divergent
denominator is annihilated by the linear DOS, $\chi _{M}\sim \int D\left(
E\right) \frac{1}{E}dE\sim finite$.

\section{Discussion}

There are several materials which possess the structures of the spin Kagom%
\'{e} lattice. Experimental data by Muon Spin Relaxation show that the
compound SrCr$_{8-x}$Ga$_{4+x}$O$_{19}$ lacks a long-range order until $0.05$%
K \cite{Uemura90prl}. The entropy measurement of the same compound gives a $%
T^{2}$ law for the specific heat at low temperatures, which was regarded as
an evidence to support the absence of long-range order \cite{Ramirez2000}.
However, spontaneous breaking of a continuous symmetry produces massless
Goldstone modes \cite{Anderson97}, and LRO can only exist at zero
temperature for 2D systems according to Mermin-Wagner theorem \cite%
{Auerbachbook}. From the present calculation of SBMFT, the special structure
of the Kagom\'{e} lattice leads to the cone structure of the spectra for the
quasi-particles in the momentum space. The existence of long-range
correlation is consistent with the picture of the gapless spectrum of
quasi-particles and the ordered ground state can produce the $T^{2}$ law of
the specific heat. Of course, the possible existence of the additional
next-nearest-neighbor coupling could further weaken the long-range
correlation.

We thank C. Broholm for helpful discussions. This work was supported by the
COE-SUG Grant (No. M58070001) of NTU and the Research Grant Council of Hong
Kong under Grant No.: HKU 703804.

\appendix{}

\section{Static spin structure factors}

The spin structure factor at zero temperature (see Fig. 5) contains both
intra-sublattice and inter-sublattice contributions,%
\begin{align}
\chi _{S}\left( \mathbf{Q}\right) & =\left\langle S_{\mathbf{Q}}^{z}S_{-%
\mathbf{Q}}^{z}\right\rangle =\frac{1}{N_{\Lambda }}\sum_{m,n\in
A,B,C}\left\langle S_{m}^{z}S_{n}^{z}\right\rangle e^{i\mathbf{Q}\cdot
\left( \mathbf{r}_{m}-\mathbf{r}_{n}\right) }  \notag \\
& =\chi _{AA}\left( \mathbf{Q}\right) +\chi _{BB}\left( \mathbf{Q}\right)
+\chi _{CC}\left( \mathbf{Q}\right)  \notag \\
& +\chi _{AB}\left( \mathbf{Q}\right) +\chi _{BA}\left( \mathbf{Q}\right)
+\chi _{AC}\left( \mathbf{Q}\right)  \notag \\
& +\chi _{CA}\left( \mathbf{Q}\right) +\chi _{BC}\left( \mathbf{Q}\right)
+\chi _{CB}\left( \mathbf{Q}\right)
\end{align}%
where we have defined the spin density wave operator%
\begin{equation}
S_{\mathbf{Q}}^{z}=\sum_{m\in A,B,C}S_{m}^{z}e^{i\mathbf{Q}\cdot \mathbf{r}%
_{m}}.  \label{spindensitywave}
\end{equation}%
The intra-sublattice and inter-sublattice contributions are%
\begin{equation*}
\chi _{AA}\left( \mathbf{Q}\right) =\int \frac{d^{2}k}{2A_{BZ}}\left[
O_{a12}\left( \mathbf{k}+\mathbf{Q}\right) +P_{a12}\left( \mathbf{k}+\mathbf{%
Q}\right) \right] Q_{a12}\left( \mathbf{k}\right) ,
\end{equation*}%
\begin{equation*}
\chi _{BB}\left( \mathbf{Q}\right) =\int \frac{d^{2}k}{2A_{BZ}}\left[
O_{a2}\left( \mathbf{k}+\mathbf{Q}\right) +P_{a13}\left( \mathbf{k}+\mathbf{Q%
}\right) \right] Q_{a13}\left( \mathbf{k}\right) ,
\end{equation*}%
\begin{equation*}
\chi _{CC}\left( \mathbf{Q}\right) =\int \frac{d^{2}k}{2A_{BZ}}\left[
O_{a1}\left( \mathbf{k}+\mathbf{Q}\right) +P_{a23}\left( \mathbf{k}+\mathbf{Q%
}\right) \right] Q_{a23}\left( \mathbf{k}\right) ,
\end{equation*}%
\begin{align*}
\chi _{AB}\left( \mathbf{Q}\right) & =\chi _{BA}\left( \mathbf{Q}\right) \\
& =\int \frac{d^{2}k}{2A_{BZ}}\left\{ \left[ -O_{b23}\left( \mathbf{k}+%
\mathbf{Q}\right) +P_{b23}\left( \mathbf{k}+\mathbf{Q}\right) \right]
Q_{b23}\left( \mathbf{k}\right) -R_{1}\left( \mathbf{k}\right) R_{1}\left(
\mathbf{k}+\mathbf{Q}\right) \right\} ,
\end{align*}%
\begin{align*}
\chi _{BC}\left( \mathbf{Q}\right) & =\chi _{CB}\left( \mathbf{Q}\right) \\
& =\int \frac{d^{2}k}{2A_{BZ}}\left\{ \left[ -O_{b13}\left( \mathbf{k}+%
\mathbf{Q}\right) +P_{b13}\left( \mathbf{k}+\mathbf{Q}\right) \right]
Q_{b13}\left( \mathbf{k}\right) -R_{2}\left( \mathbf{k}\right) R_{2}\left(
\mathbf{k}+\mathbf{Q}\right) \right\} ,
\end{align*}%
\begin{align*}
\chi _{CA}\left( \mathbf{Q}\right) & =\chi _{AC}\left( \mathbf{Q}\right) \\
& =\int \frac{d^{2}k}{2A_{BZ}}\left\{ \left[ -O_{b12}\left( \mathbf{k}+%
\mathbf{Q}\right) +P_{b12}\left( \mathbf{k}+\mathbf{Q}\right) \right]
Q_{b12}\left( \mathbf{k}\right) -R_{3}\left( \mathbf{k}\right) R_{3}\left(
\mathbf{k}+\mathbf{Q}\right) \right\} ,
\end{align*}%
where
\begin{subequations}
\begin{equation*}
O_{a1}\left( \mathbf{k}\right) =\frac{\Delta ^{2}\gamma _{1}^{2}\left(
\mathbf{k}\right) }{1-\omega ^{2}\left( \mathbf{k}\right) },O_{a2}\left(
\mathbf{k}\right) =\frac{\Delta ^{2}\gamma _{2}^{2}\left( \mathbf{k}\right)
}{1-\omega ^{2}\left( \mathbf{k}\right) },O_{a3}\left( \mathbf{k}\right) =%
\frac{\Delta ^{2}\gamma _{3}^{2}\left( \mathbf{k}\right) }{1-\omega
^{2}\left( \mathbf{k}\right) },
\end{equation*}%
\end{subequations}
\begin{subequations}
\begin{equation*}
P_{a12}\left( \mathbf{k}\right) =\frac{\Delta ^{2}\left[ \gamma
_{1}^{2}\left( \mathbf{k}\right) +\gamma _{2}^{2}\left( \mathbf{k}\right) %
\right] }{2\omega \left( \mathbf{k}\right) \left[ 1-\omega \left( \mathbf{k}%
\right) \right] },P_{a13}\left( \mathbf{k}\right) =\frac{\Delta ^{2}\left[
\gamma _{1}^{2}\left( \mathbf{k}\right) +\gamma _{3}^{2}\left( \mathbf{k}%
\right) \right] }{2\omega \left( \mathbf{k}\right) \left[ 1-\omega \left(
\mathbf{k}\right) \right] },P_{a23}\left( \mathbf{k}\right) =\frac{\Delta
^{2}\left[ \gamma _{3}^{2}\left( \mathbf{k}\right) +\gamma _{2}^{2}\left(
\mathbf{k}\right) \right] }{2\omega \left( \mathbf{k}\right) \left[ 1-\omega
\left( \mathbf{k}\right) \right] },
\end{equation*}%
\end{subequations}
\begin{subequations}
\begin{equation*}
Q_{a12}\left( \mathbf{k}\right) =\frac{\Delta ^{2}\left[ \gamma
_{1}^{2}\left( \mathbf{k}\right) +\gamma _{2}^{2}\left( \mathbf{k}\right) %
\right] }{2\omega \left( \mathbf{k}\right) \left[ 1+\omega \left( \mathbf{k}%
\right) \right] },Q_{a13}\left( \mathbf{k}\right) =\frac{\Delta ^{2}\left[
\gamma _{1}^{2}\left( \mathbf{k}\right) +\gamma _{3}^{2}\left( \mathbf{k}%
\right) \right] }{2\omega \left( \mathbf{k}\right) \left[ 1+\omega \left(
\mathbf{k}\right) \right] },Q_{a23}\left( \mathbf{k}\right) =\frac{\Delta
^{2}\left[ \gamma _{3}^{2}\left( \mathbf{k}\right) +\gamma _{2}^{2}\left(
\mathbf{k}\right) \right] }{2\omega \left( \mathbf{k}\right) \left[ 1+\omega
\left( \mathbf{k}\right) \right] },
\end{equation*}%
\end{subequations}
\begin{subequations}
\begin{equation*}
O_{b23}\left( \mathbf{k}\right) =\frac{\Delta ^{2}\gamma _{3}\left( \mathbf{k%
}\right) \gamma _{2}\left( \mathbf{k}\right) }{1-\omega ^{2}\left( \mathbf{k}%
\right) },O_{b13}\left( \mathbf{k}\right) =\frac{\Delta ^{2}\gamma
_{1}\left( \mathbf{k}\right) \gamma _{3}\left( \mathbf{k}\right) }{1-\omega
^{2}\left( \mathbf{k}\right) },O_{b12}\left( \mathbf{k}\right) =\frac{\Delta
^{2}\gamma _{1}\left( \mathbf{k}\right) \gamma _{2}\left( \mathbf{k}\right)
}{1-\omega ^{2}\left( \mathbf{k}\right) },
\end{equation*}%
\end{subequations}
\begin{subequations}
\begin{equation*}
P_{b23}\left( \mathbf{k}\right) =\frac{\Delta ^{2}\gamma _{3}\left( \mathbf{k%
}\right) \gamma _{2}\left( \mathbf{k}\right) }{2\omega \left( \mathbf{k}%
\right) \left[ 1-\omega \left( \mathbf{k}\right) \right] },P_{b13}\left(
\mathbf{k}\right) =\frac{\Delta ^{2}\gamma _{1}\left( \mathbf{k}\right)
\gamma _{3}\left( \mathbf{k}\right) }{2\omega \left( \mathbf{k}\right) \left[
1-\omega \left( \mathbf{k}\right) \right] },P_{b12}\left( \mathbf{k}\right) =%
\frac{\Delta ^{2}\gamma _{1}\left( \mathbf{k}\right) \gamma _{2}\left(
\mathbf{k}\right) }{2\omega \left( \mathbf{k}\right) \left[ 1-\omega \left(
\mathbf{k}\right) \right] },
\end{equation*}%
\end{subequations}
\begin{subequations}
\begin{equation*}
Q_{b23}\left( \mathbf{k}\right) =\frac{\Delta ^{2}\gamma _{3}\left( \mathbf{k%
}\right) \gamma _{2}\left( \mathbf{k}\right) }{2\omega \left( \mathbf{k}%
\right) \left[ 1+\omega \left( \mathbf{k}\right) \right] },Q_{b13}\left(
\mathbf{k}\right) =\frac{\Delta ^{2}\gamma _{1}\left( \mathbf{k}\right)
\gamma _{3}\left( \mathbf{k}\right) }{2\omega \left( \mathbf{k}\right) \left[
1+\omega \left( \mathbf{k}\right) \right] },Q_{b12}\left( \mathbf{k}\right) =%
\frac{\Delta ^{2}\gamma _{1}\left( \mathbf{k}\right) \gamma _{2}\left(
\mathbf{k}\right) }{2\omega \left( \mathbf{k}\right) \left[ 1+\omega \left(
\mathbf{k}\right) \right] },
\end{equation*}%
\end{subequations}
\begin{subequations}
\begin{equation*}
R_{1}\left( \mathbf{k}\right) =\frac{\Delta \gamma _{1}\left( \mathbf{k}%
\right) }{2\omega \left( \mathbf{k}\right) },R_{2}\left( \mathbf{k}\right) =%
\frac{\Delta \gamma _{2}\left( \mathbf{k}\right) }{2\omega \left( \mathbf{k}%
\right) },R_{3}\left( \mathbf{k}\right) =\frac{\Delta \gamma _{3}\left(
\mathbf{k}\right) }{2\omega \left( \mathbf{k}\right) }.
\end{equation*}

\end{subequations}


\begin{thebibliography}{99}
\bibitem{Anderson} P. W. Anderson, Mater. Res. Bull. \textbf{8}, 153 (1973);
P. Fazekas and P. W. Anderson, Philos. Mag. \textbf{30}, 423 (1974).

\bibitem{Huse} D. A. Huse and V. Elser, Phys. Rev. Lett. \textbf{60}, 2531
(1988).

\bibitem{Jolicoeur} Th. Jolicoeur and J. C. Le Guillou, Phys. Rev. B \textbf{%
40}, 2727 (1989).

\bibitem{Bernu} B. Bernu, C. Lhuillier, and L. Pierre, Phys. Rev. Lett.
\textbf{69}, 2590 (1992); B. Bernu, P. Lecheminant, C. Lhuillier, and L.
Pierre, Phys. Rev. B \textbf{50}, 10048 (1994).

\bibitem{Capriotti} L. Capriotti, A. E. Trumper, and S. Sorella, Phys. Rev.
Lett. \textbf{82}, 3899 (1999).

\bibitem{Sachdev} S. Sachdev, Phys. Rev. B \textbf{45}, 12377 (1992).

\bibitem{Misguich} See a review in G. Misguich and C. Lhuillier, \textit{%
Frustrated Spin Systems}, edited by H. T. Diep, (World-Scientific,
Singapore, 2004).

\bibitem{Harris} A. B. Harris, C. Kallin, and A. J. Berlinsky, Phys. Rev. B
\textbf{45}, 2899 (1992).

\bibitem{Zeng} C. Zeng and V. Elser, Phys. Rev. B \textbf{42}, 8436 (1990).

\bibitem{Chalker} J. T. Chalker and J. F G. Eastmond, Phys. Rev. B \textbf{46%
}, 14201 (1992).

\bibitem{Waldtmann} C. Waldtmann, H. -U. Everts, B. Bernu, C. Lhuillier, P.
Sindzingre, P. Lecheminant, L. Pierre, Eur. Phys. J. B \textbf{2}, 501
(1998).

\bibitem{Syromyatnikov} A. V. Syromyatnikov and S. V. Maleyev, Phys. Rev. B
\textbf{66}, 132408 (2002); P. Nikolic and T. Senthil, Phys. Rev. B \textbf{%
68}, 214415 (2003).

\bibitem{Broholm} C. Broholm, G. Aeppli, G. P. Espinosa, and A. S. Cooper,
Phys. Rev. Lett. \textbf{65}, 3173 (1990); A. Keren \textit{et al}., Phys.
Rev. B \textbf{53}, 6451 (1996); A. Fukaya \textit{et al}., Phys. Rev. Lett.
\textbf{91}, 207603 (2003).

\bibitem{Ramirez1990} A. P. Ramirez, G. P. Espinosa, and A. S. Cooper, Phys.
Rev. Lett. \textbf{64}, 2070 (1990); B. Martinez \textit{et al.}, Phys. Rev.
B \textbf{46}, 10786 (1992).

\bibitem{Ramirez2000} A. P. Ramirez, B. Hessen, and M. Winklemann, Phys.
Rev. Lett. \textbf{84}, 2957 (2000).

\bibitem{Sindzingre} P. Sindzingre, G. Misguich, C. Lhuillier, B. Bernu, L.
Pierre, C. Waldtmann, and H. -U. Everts, Phys. Rev. Lett. \textbf{84}, 2953
(2000).

\bibitem{Budnik} R. Budnik and Assa Auerbach, Phys. Rev. Lett. \textbf{93},
187205 (2004).

\bibitem{Ran} Y. Ran, M. Hermele, P. A. Lee, and X. G. Wen, cond-mat/061141.

\bibitem{Anderson87} P. W. Anderson, Science \textbf{235}, 1196 (1987).

\bibitem{Auerbach} D. P. Arovas and A. Auerbach, Phys. Rev. B \textbf{38},
316 (1988); A. Auerbach and D. P. Arovas, Phys. Rev. Lett. \textbf{61}, 617
(1988).

\bibitem{Auerbachbook} A. Auerbach, \textit{Interacting Electrons and
Quantum Magnetism}, (Springer-Verlag, NY, 1994).

\bibitem{Gazza} C. J. Gazza and H. A. Ceccatto, J. Phys.: Condens. Matter
\textbf{5}, L135 (1993); L. O. Manuel, A. E. Trumper, and H. A. Ceccatto,
Phys. Rev. B \textbf{57}, 8348 (1998).

\bibitem{Shen} S. -Q. Shen and F. C. Zhang, Phys. Rev. B \textbf{66}, 172407
(2002).

\bibitem{Manuel} L. O. Manuel, A. E. Trumper, C. J. Gazza, and H. A.
Ceccatto, Phys. Rev. B \textbf{50}, 1313 (1994).

\bibitem{Wang} F. Wang and A. Vishwanath, Phys. Rev. B \textbf{74}, 174423
(2006).

\bibitem{Read} N. Read and D. M. Newns, J. Phys. C \textbf{16}, 3273 (1983).

\bibitem{Uemura90prl} Y. J. Uemura, A. Keren, K. Kojima, L. P. Le, G. M.
Luke, W. D. Wu, Y. Ajiro, T. Asano, Y. Kuriyama, M. Mekata, H. Kikuchi, and
K. Kakurai, Phys. Rev. Lett. \textbf{73}, 3306 (1994)

\bibitem{Anderson97} P. W. Anderson, \textit{Concepts in Solids}, (World
Scientific, Singapore, 1997).
\end{thebibliography}
\end{document}